\documentclass[11pt]{article}

\usepackage{graphicx}
\usepackage{amsmath}
\usepackage{amssymb}
\usepackage{enumerate}
\usepackage{xcolor}
\usepackage{bm}
\usepackage[T1]{fontenc}
\usepackage{float}
\usepackage{soul} 
\usepackage{multirow}
\usepackage{comment}
\usepackage{moreverb,url}
\usepackage{setspace}

\usepackage{booktabs}
\usepackage{tabularx}

\newcommand\clearrow{\global\let\rowmac\relax}
\clearrow

\usepackage{natbib}
\usepackage{amsthm,amssymb,amsmath}

\newtheorem{theorem}{Theorem}

\def \bY { \boldsymbol Y }

\def \bt { \boldsymbol t }

\def \bV { \boldsymbol V }

\def \bb { \boldsymbol b }

\def \bu { \boldsymbol u }

\def \bbeta { \boldsymbol \beta }
\def \balpha { \boldsymbol \alpha }

\def \btheta { \boldsymbol \theta }
\def \bTheta { \boldsymbol \Theta }

\def \tT { \tilde{T} }

\def\bSig\mathbf{\Sigma}

\title{\textbf{Multi-Layer Backward Joint Model for Dynamic Prediction of Clinical Events with Multivariate Longitudinal Predictors of Mixed Types}}

\date{}
\begin{document}

\maketitle

\begin{center}
\author{Wenhao Li$^{1}$, Zhe Yin$^{1}$, Liang Li$^{1,*}$\\
  $^{1}$ Department of Biostatistics, The University of Texas MD Anderson Cancer Center, Houston, TX, U.S.A. \\
  $^{*}$ \emph{email}: LLi15@mdanderson.org }
\end{center}

\begin{abstract}
	Dynamic prediction of time-to-event outcomes using longitudinal data is highly useful in clinical research and practice. A common strategy is the joint modeling of longitudinal and time-to-event data. The shared random effect model has been widely studied for this purpose. However, it can be computationally challenging when applied to problems with a large number of longitudinal predictor variables, particularly when mixed types of continuous and categorical variables are involved. Addressing these limitations, we introduce a novel multi-layer backward joint model (MBJM). The model structure consists of multiple data layers cohesively integrated through a series of conditional distributions that involve longitudinal and time-to-event data, where the time to the clinical event is the conditioning variable. This model can be estimated with standard statistical software with rapid and robust computation, regardless of the dimension of the longitudinal predictor variables. We provide both theoretical and empirical results to show that the MBJM outperforms the static prediction model that does not fully account for the longitudinal nature of the prediction. In an empirical comparison with the shared random effects joint model, the MBJM demonstrated competitive performance with substantially faster and more robust computation. Both the simulation and real data application from a primary biliary cirrhosis study utilized seven longitudinal biomarkers, five continuous and two categorical, larger than the typically published joint modeling problems.\\
\end{abstract}

\noindent%
\vfill

\section{Introduction}
\label{sec:intro}
In a typical longitudinal study in population health, a cohort of individuals are enrolled and their longitudinal data are measured at clinical visits until the end of the follow-up, which could be either a censoring time or the occurrence of a terminal adverse event of scientific or clinical interest. The longitudinal data may include lab tests of biomarkers, quality of life or other behavioral and social environmental measures, numerical summaries of the clinical history, etc. Since these data are quantification of an individual's health status over time, we refer to them all as ``longitudinal biomarker data'' in this paper. The longitudinal biomarkers are often associated with the terminal adverse event, and hence can serve as the foundation for statistical prediction models of an individual's risk of experiencing that event. The desirable prognostic tool would generate predicted probabilities at each clinical visit, taking into account the evolving longitudinal history up to the time of prediction. This type of problem is commonly referred to as dynamic prediction (DP). \citep{Taylor2005, Yu2008, Rizopoulos2011, Taylor2013} Two mainstream approaches to DP exist: joint modeling (JM) and landmark modeling (LM). \citep{Rizopoulos2012, Zheng2005, vanHouwelingen2011} Each has its advantages and disadvantages. \citep{Li2016, Li2023} The primary focus of this paper is on the JM method.

The JM estimates the joint distribution of longitudinal and survival data. Once this distribution is determined, the risk of a future event for a patient at risk by the prediction time can be easily calculated as a conditional probability. \citep{Rizopoulos2011, EchouffoTcheugui2012, Andrinopoulou2021, Shen2021} As in the usual prediction model problems, augmenting the number of longitudinal predictors can often enhance the prediction accuracy. Nevertheless, this may impose a heavy computational burden on many JM methods. The log-likelihood function of the widely-used shared random effects model cannot be solved analytically and often necessitates numerical integration with respect to the random effects, followed by optimization of the log-likelihood function, which is not guaranteed to be concave. The dimension of the integral escalates with an increasing number of longitudinal biomarkers, potentially leading to numerical difficulties. \citep{Rizopoulos2012, Elashoff2016, Hickey2018} Recently, a novel backward joint model (BJM) has been proposed for DP. \citep{Shen2021, Li2023b} By decomposing the likelihood of the joint distribution into the distribution of time-to-event and the distribution of longitudinal data conditional on time-to-event, the BJM avoids using the random effects and can effectively handle a large number of longitudinal biomarkers with stable and tractable computation.  

In the JM literature, models with categorical longitudinal data have received less attention.\citep{Choi2013} Categorical longitudinal data have less granularity of information than continuous longitudinal data and may increase numerical instability in model fitting. Limited literature exists on the inclusion of categorical biomarkers in a joint model designed for dynamic prediction problems, \citep{li2010joint, li2012joint, Andrinopoulou2014, Andrinopoulou2015} and most of these studies only considered one or two longitudinal biomarkers. In this paper, we study a data application with a substantially larger number of categorical or continuous biomarkers (2 categorical and 5 continuous). This is accomplished by an extension of the previous BJM framework. This extension is not trivial. The hallmark feature of the BJM framework is that the likelihood has, at most, one-dimensional numerical integrals in it, regardless of how many longitudinal variables are used. This desirable computational advantage is lost if we simply replace the linear mixed model for continuous longitudinal data with a generalized linear mixed model in the presence of categorical data in the BJM. Therefore, we propose a novel multi-layer backward joint model (MBJM) that comprises multiple layers stacked on top of each other in the likelihood factorization, each corresponding to a longitudinal biomarker variable. Each layer can be estimated separately using a linear mixed model or generalized linear mixed model, available from standard software packages. This multi-layered structure improves the flexibility and robustness of the joint model. This idea resembles that of the deep neural network, which consists of stacked layers formulated by simple models. This paper presents the model and its motivation in Section 2, the estimation procedure in Section 3, some theoretical results in Section 4, a real data application in Section 5, comparative simulation studies in Section 6, and a comprehensive discussion of the proposed methodology in Section 7.

\section{Estimand and Model}
\label{sec:estimand_model}

\subsection{Data and notation}
Let the subjects in the training dataset be indexed by $i = 1, 2,...,n$. Each subject has $M$ longitudinal biomarkers, and subject $i$'s longitudinal measurements are denoted by $M$ vectors $\bm{Y}_{1i}, \bm{Y}_{2i},...,\bm{Y}_{Mi}$. These can be either continuous or categorical biomarkers. For the $m$-th biomarker, the vector $\bY_{mi}$ includes $n_{mi}$ repeated measurements. The $j$-th measurement is denoted by $Y_{mij}$, where $j = 1,2,...,n_{mi}$, and the corresponding measurement time is denoted by $t_{mij}$. The baseline covariates are collected in the vector $\bV_i$. We denote the time of the outcome event of interest by $\tT_i$ and the censoring time by $C_i$. We make the common assumption in joint modeling literature that the censoring time $C_i$ is conditionally independent of $\tT_i$ and $\bY_i$ given $\bV_i$, \citep{WulfsohnTsiatis1997, Handbook.Longitudinal.Chap15, tsiatis2004joint} and discuss dealing with deviation from this assumption in Section \ref{sec:disc}. The observed event time is $T_i = \min(\tT_i, C_i)$ and the censoring indicator $\delta_i = \bm{1}(\tT_i < C_i)$. As in most joint modeling literature, we assume that the biomarker measurement times are non-informative, i.e., distribution of $\{ t_{mij} \}$ in $(0, T_i)$ does not depend on $T_i, C_i, \bV_i$ or the longitudinal biomarker data. 

\subsection{The estimand for prediction}
\label{sec:estimand}
The objective of this research is to predict survival probability dynamically. We use subscript $o$ to denote a new generic subject for whom the prediction will be made. This subject's data are independently and identically distributed as the subjects in the training dataset. The prediction time is denoted by $s$ and the pre-specified prediction horizon is $\Delta$. We use $\overline{\bm{Y}_o(s)}$ to denote the history of all the longitudinal biomarkers before time $s$, including both continuous and categorical biomarkers. The dynamic prediction of the risk probability is given by : 
	\begin{eqnarray}\label{DP_risk}
		 P(s<\tilde{T}_o \leq s +\Delta | \overline{\bm{Y}_o(s)}, \tT_o > s, C_o > s, \bV_o)  =  \dfrac{P( \overline{\bm{Y}_o(s)},  s<\tilde{T}_o \leq s +\Delta |\bV_o) }{P( \overline{\bm{Y}_o(s)}, \tT_o > s| \bV_o)}.
	\end{eqnarray}
This is the probability of experiencing an event within the time interval of $s$ and $s + \Delta$, given that the subject is still at risk ($\tT_o > s$) and under follow-up ($C_o > s$) at the prediction time $s$, and using the observed history $\{ \overline{\bm{Y}_o(s)}, \bV_i \}$. As (\ref{DP_risk}) suggests, this probability can be calculated directly from the conditional joint distribution of survival and longitudinal data given the baseline predictors, i.e., $P( \tT_o , \overline{\bm{Y}_o( \tT_o )} | \bV_o ) $, which can be learned from the training data. Therefore, we will focus on the estimation of this conditional joint distribution in subsequent sections.

\subsection{The multi-layer backward joint model}
\label{sec:MBJM}

We designate the vector $\bY_o =  ( \bm{Y}_{1o}^T, \bm{Y}_{2o}^T,...,\bm{Y}_{Mo}^T )^T $ to represent all the longitudinal biomarker measurements of the generic subject $o$. In order to estimate the conditional joint distribution $P(\bY_o, \tT_o | \bV_o)$, we decompose it into two components: $P(\bY_o, \tT_o | \bV_o) = P(\bY_o | \tT_o, \bV_o) P(\tT_o | \bV_o)$.
This expression suggests two indispensable models: the survival sub-model $P(\tT_o | \bV_o)$ and the longitudinal sub-model $P(\bm{Y}_o | \tT_o, \bV_o)$. The former can be formulated by any appropriate regression model of time-to-event data using subject-specific (i.e., time-invariant) covariates, such as the Weibull model, Cox model, accelerated failure time model and their extensions. The latter can be formulated by any appropriate regression model of multivariate longitudinal data using subject-specific covariates. When some or all of $\bm{Y}_o$ are categorical variables, the latter model must include generalized linear mixed model or its extensions, for which the likelihood function does not have a closed-form expression. \citep{DiscreteLongitudinalBook} Numerical integration with respect to random effects must be used to account for the intrasubject correlations among repeated measurements within the same longitudinal biomarker and between different longitudinal biomarkers. The dimension of the integration increases with the number of categorical longitudinal biomarkers and the complexity of the correlation structure. This difficulty does not exist when all of  $\bm{Y}_o$ are continuous variables, \citep{Shen2021, Li2023b} because the random effects in typical multivariate linear mixed models can be integrated out, yielding a closed-form likelihood function without numerical integration. This paper proposes an alternative formulation of $P(\bm{Y}_o | \tT_o, \bV_o)$ to avoid this difficulty. Our goal is to produce a computationally efficient and robust algorithm that can be implemented quickly by data analysts with standard software. 

\begin{figure}[H]
    \centering
    \includegraphics[width = 1\textwidth]{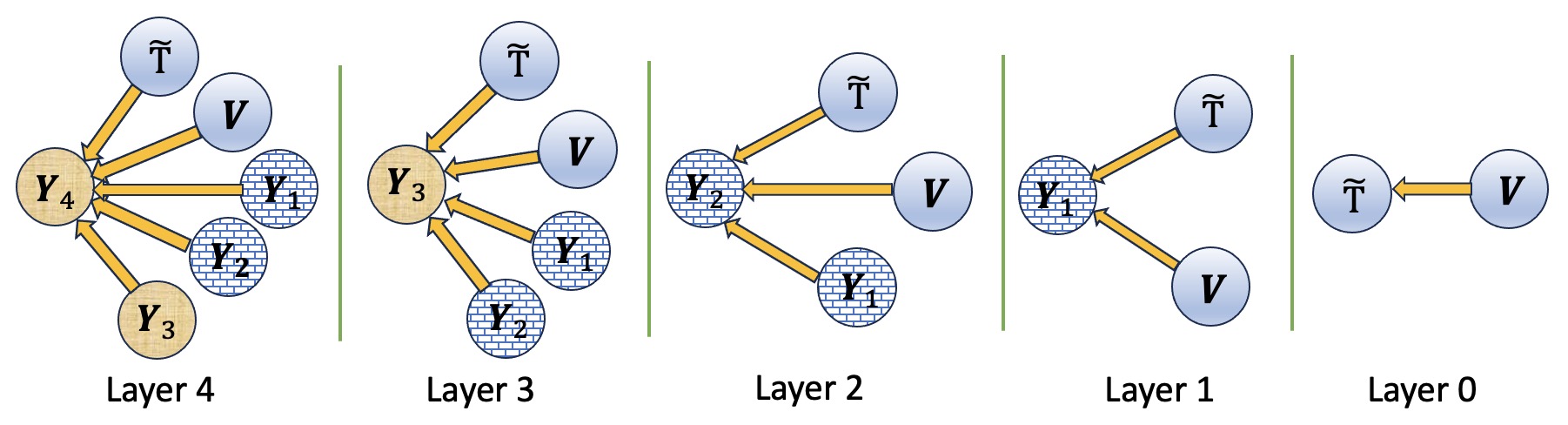}
    \caption{ \textit{An illustration of an MBJM. $\tT$ denotes the survival time. $\bV$ denotes baseline covariates. This example involves four longitudinal biomarkers: $\bY_1$ and $\bY_2$ are categorical, and $\bY_3$ and $\bY_4$ are continuous. Layer 0 corresponds to using the baseline covariates to model the survival time. Moving forward, Layer 1 models the first categorical biomarker given $\tT$ and $\bV$; Layer 2 models the second categorical biomarker given $\bY_1$, $\tT$, and $\bV$, etc. The categorical biomarkers are modeled in lower layers because there are fewer model parameters in these layers and categorical outcomes have less granularity of information than continuous outcomes. } }  \label{MBJM_model}
\end{figure}

To accommodate the correlation among longitudinal biomarker data, the proposed algorithm organizes different biomarkers into distinct layers. Without loss of generality, let $\bm{Y}_{1o}, \bm{Y}_{2o},..,\bm{Y}_{Mo}$ to represent the first, second, ..., and $M$-th layer, respectively. The distribution of the $m$-th biomarker $\bm{Y}_{mo}$ is modeled conditional on biomarkers from previous layers, i.e., $\bm{Y}_{1o}, \bm{Y}_{2o},..,\bm{Y}_{m-1o}$, as well as the time-invariant covariates $\bV_o$ and $\tT_o$. Hence, the joint distribution of $\bm{Y}_{1o}, \bm{Y}_{2o},..,\bm{Y}_{Mo}$ can be factorized as
 \begin{eqnarray}
 P(\bm{Y}_o | \tT_o,  \bV_o) = \prod_{m = 1}^{M} P(\bm{Y}_{mo}| 
 \bm{Y}_{1o}, \bm{Y}_{2o},..,\bm{Y}_{m-1o},\tT_o,  \bV_o).  
 \label{eq:series.conditional}
 \end{eqnarray}
Figure \ref{MBJM_model} illustrates the structure of this formulation.

Equation (\ref{DP_risk}) signifies that there exists an integral of $\tT_o$ from $s$ to infinity, which is  $ P( \overline{\bm{Y}_o(s)}, \tT_o > t | \bV_o ) = \int_t^{\infty} P( \overline{\bm{Y}_o(s)}, \tT_o = u | \bV_o ) du, ~(t \geqslant s) $. This integral can be calculated under the assumption that the support of $\tT_o$ is contained within the support of $C_o$, conditional on $\bV_o$. Supporting evidence for this assumption is that the Kaplan-Meier curve of $\tT_o$ descends to 0. When this assumption is violated, $P( \overline{\bm{Y}_o(s)}, \tT_o = u | \bV_o )$ cannot be estimated at every $u$ in the interval $(t, \infty)$ and the integral above is undefined. Two methods can address this issue. \citep{Li2023b} The first is extrapolation (MBJM-EX), which assumes that the estimated distribution function of $\tT_o | \bV_o $ extends parametrically beyond the range of the observed survival data. This approach is practical when there is only moderate violation of the assumption on support. Another approach is the two-part model (MBJM-TP). The two-part model necessitates an additional longitudinal submodel for individuals who survive beyond a pre-specified maximum follow-up time $\tau_{max}$: $\tT > \tau_{max}$. These individuals are referred to as long-term survivors (LTS). \citep{SKWang2020Biostat, Shen2021} The other individuals who had the event before $\tau_{max}$ are called non-LTS. $ P( \overline{\bm{Y}_o(s)}, \tT_o > t | \bV_o )$ can be rewritten as $\int_t^{\tau_{max}} P( \overline{\bm{Y}_o(s)} | \tT_o = u, \bV_o ) P( \tT_o = u | \bV_o ) du + P( \overline{\bm{Y}_o(s)} | \tT_o > \tau_{max}, \bV_o ) P( \tT_o > \tau_{max} | \bV_o ) $. The first term can be estimated by the longitudinal submodel for non-LTS and the second term by LTS. This approach works when there is severe violation of the assumption on support, at the expense of added model parameters. Further insight into the extrapolation and two-part model techniques applied in the backward joint model can be found in Li et al. \cite{Li2023b} 


\subsection{Two sub-models}
\label{sec:two-part}

The estimation of the longitudinal sub-model $P(\bm{Y}_o | \tT_o, \bV_o)$ is based on linear mixed model or generalized linear mixed model. The covariates may include $\tT_i, \bV_i, \bt_i$ and their interactions as fixed effects. Additionally, random intercept and slope terms are included as random effects. The interpretation of these models is that conditional on the stratification variable $\tT_i$ (i.e., among individuals with the same survival time, or equivalently, length of longitudinal biomarker trajectories), how the biomarker trajectories are affected by $\bV_i$, $\bt_i$ and random effects. The way $\tT_i$ enters these models reflects the data analyst's assumption that the longitudinal biomarker trajectories vary with the stratification variable $\tT_i$. It is advantageous to use a flexible model specification, especially regarding the effect of $\tT_i$. 

We consider subject $i$'s $m$-th biomarker measured at the $j$-th visit. If this biomarker is continuous, it can be modeled using a linear mixed model:
\begin{eqnarray}\label{BJMmodel_con}
 &&Y_{mij} =  \beta_{m0} +  t_{mij}\beta_{m1} + \bV_{i}^T \bbeta_{m2}  +  Y_{1ij}  \nu_{m1}\nonumber +\ Y_{2ij}\nu_{m2} +... + Y_{m-1ij}  \nu_{m(m - 1)} + \nonumber \\
 && ~~~~~~~~~ g(\tT_i) \beta_{m3}  +  \bu_{mij}^T \bb_{mi} + \epsilon_{mij}. 
\end{eqnarray} 
If this biomarker is categorical, it can be modeled using a generalized linear mixed model:
\begin{eqnarray}\label{BJMmodel_cate}
&&P(Y_{mij} = 1 | Y_{1ij}, Y_{2ij},...,Y_{(m-1)ij}, \tT_i, \bV_{i}, \bb_{mi}) =  \mathrm{expit}( \beta_{m0} +  t_{mij}\beta_{m1} + \bV_{i}^T \bbeta_{m2}    \nonumber \\
&& +\  Y_{1ij}\nu_{m1} +  Y_{2ij}\nu_{m2} +... + Y_{(m-1)ij}  \nu_{m(m - 1)} + g(\tT_i) \beta_{m3}  +  \bu_{mij}^T \bb_{mi}), 
\end{eqnarray} 
where $\mathrm{expit}(x) = \exp(x)/\{1 + \exp(x)\}$. The function $g(\cdot)$ denotes a pre-specified transformation applied to the original true survival time $\tT_i$ to make the linearly additive assumption of these models more plausible. Common choices are  logarithm or square root transformation, which mitigate the skewness in survival time. The residuals $\varepsilon_{mi}(t)$ are independent random errors with a mean of zero and a variance of $\sigma_{e,m}^2$. We have the flexibility to incorporate various types of random effects, such as the random intercept ($\bu_{mij} = 1$), a combination of random intercept and slope ($\bu_{mij} = (1, t_{mij})^T$), or nonlinear random effect terms ($\bu_{mij}$ includes additional spline basis terms or polynomial terms). The other biomarkers \{$Y_{1ij}$, $Y_{2ij}$,...,$Y_{m-1ij}$\} are considered as fixed effects in the longitudinal sub-model for $Y_{mij}$. The random effects associated with all $M$ biomarkers, denoted as $\bm{b}_{i} =(\bm{b}_{1i}, \bm{b}_{2i},...,\bm{b}_{Mi})^T$, are assumed to follow a multivariate normal distribution with mean zero and covariance matrix $\bm{\Omega}$. When the longitudinal sub-model model includes random intercepts and slopes, $\bm{\Omega}$ is a $2M \times 2M$ matrix; when it includes only the random intercepts, $\bm{\Omega}$ is $M \times M$. The $\bm{\Omega}$ is a block diagonal matrix, where the random effects associated with a biomarker are correlated, but the random effects associated with different biomarkers are independent. This simplification is plausible because the longitudinal mixed models for the right-hand side of (\ref{eq:series.conditional}) are independent of each other. The random effects account for the longitudinal correlation of a biomarker's repeated measures. The correlation among different biomarkers is accounted for by including  $\{Y_{1ij}, Y_{2ij},...,Y_{(m-1)ij}\}$ as fixed effects in the regression model for $Y_{mij}$. As a result, the multi-layered longitudinal sub-models can be estimated one layer at a time, with standard software for linear mixed model or generalized linear mixed model. This makes it convenient to implement this method. 

We now consider the additional longitudinal sub-model for LTS when a two-part BJM is used. If the $m$-th biomarker is continuous, 
\begin{eqnarray}\label{LTSmodel_con}
 &&Y_{mij} = \beta_{m0}^e +  t_{mij}\beta_{m1}^e + \bV_{i}^T \bbeta_{m2}^e  +  Y_{1ij}  \nu_{m1}^e +\ Y_{2ij}\nu_{m2}^e +... + Y_{m-1ij}  \nu_{m(m - 1)}^e \nonumber \\
 &&  ~~~~~~~~~~ +  \bu_{mij}^T \bb_{mi}^e + \epsilon_{mij}^e,  ~~ \tT_i > \tau_{max}. 
\end{eqnarray} 
If the $m$-th biomarker is categorical, 
\begin{eqnarray}\label{LTSmodel_cate}
&&P(Y_{mij} = 1 |Y_{1ij}, Y_{2ij},...,Y_{m-1ij}, \tT_i, V_{i}, \bb_{mi}^e) =  \mathrm{expit}( \beta_{m0}^e +  t_{mij}\beta_{m1}^e + V_{i} \beta_{m2}^e + Y_{1ij}\nu_{m1}^e   \nonumber \\
&& ~~~~~~~~~~~~~~ +\ Y_{2ij}\nu_{m2}^e +... + Y_{m-1ij}  \nu_{m(m - 1)}^e  +  \bu_{mij}^T \bb_{mi}^e), ~~~ \tT_i > \tau_{max}. 
\end{eqnarray} 
The superscript $^e$ stands for ``extra'', which signifies the inclusion of an extra longitudinal sub-model for the LTS. Both LTS models (\ref{LTSmodel_con}) and (\ref{LTSmodel_cate}) share a similar formulation as (\ref{BJMmodel_con}) and (\ref{BJMmodel_cate}), except that $\tT$ is not a covariate because all subjects with $\tT > \tau_{max}$ are modeled together. The actual survival time $\tT$ varies among the LTS subjects and is not identifiable from the data. This heterogeneity within the LTS may cause its variance-covariance matrix of random effects and residuals to be larger than those in the non-LTS models (\ref{BJMmodel_con}) and (\ref{BJMmodel_cate}). Hence, we allow the LTS models to have their own distinct parameters. 


\subsection{The static prediction model}
An important competitor of the dynamic prediction model (DPM) is the static prediction model (SPM). \citep{Li2016, Parast2019, Shen2021, Yao2023} Most of the published prediction models of clinical events can be viewed as an SPM. In subsequent sections, we present both theoretical and empirical studies to demonstrate that the DPM performs equivalently or better than the SPM. The theoretical results are general and apply to any DPM; the empirical results in this paper pertain to the MBJM only. These results highlight the importance of studying dynamic prediction problems, to which the proposed MBJM is one of the many available approaches.  

The SPM fits the regression model $\tilde{T}$ given $\bV$ to the training data. Here $\bV$ represents the predictor variables measured at baseline and $\tilde{T}$ is the time gap from baseline to the clinical event. Denote the estimated conditional distribution by $\hat{P}( \tilde{T} \leqslant \Delta | \bV = \boldsymbol{v} ) = m( \Delta, \boldsymbol{v}; \hat{\btheta})$, where $\hat{\btheta}$ is the estimated parameters. A generic subject $o$ in the validation data is at risk at the prediction time $s$. We denote by $\bV_s$ the same predictor variables as $\bV$ in the training data but measured at time $s$. The SPM defines the predicted probability as $m( \Delta, \bV_s; \hat{\btheta})$. In other words, the SPM trains the model using baseline population and baseline measurements of the predictor variables, but applies it to subjects during follow-up, ignoring the time-varying at-risk population, possibly time-varying association between the predictor and clinical event, and the longitudinal history of the predictor variables other than the ``current'' measurements of those variables at the prediction time.

\section{Estimation}
Previous research on the BJM has demonstrated that this type of models can be estimated by a complete case analysis (CCA), ignoring the censored subjects when estimating the longitudinal sub-model for non-LTS. \citep{Li2018JASA, SKWang2020Biostat, Shen2021, ShikunAOAS2023, Li2023b, Wang2023JASA} While the likelihood-based approach that uses all data can improve the statistical efficiency, the improvement is modest. \citep{SKWang2020Biostat, ShikunAOAS2023} This efficiency gain comes with more complicated algorithm; rather than relying on standard software, it needs customized computing programs tailored for different model formulations and variants. Therefore, we decided to use the CCA to fit the MBJM in this paper and hope that this approach would increase the uptake of MBJM in statistical practice. 

The BJMs are usually estimated through a two-step algorithm: a regression model for the survival time using baseline covariates, followed by a regression model for all the longitudinal data given the survival time and baseline covariates. With the MBJM, the second step above is further divided into multiple steps that correspond to the layers, and the model of each layer can be estimated independently, one after another. More specifically, we first fit the baseline survival model $P(\tT_o | \bV_o)$ using any survival regression model suitable for the data. Then we fit models (\ref{BJMmodel_con}) and (\ref{BJMmodel_cate}) for $m=1,2,..., M$ sequentially. Models (\ref{LTSmodel_con}) and (\ref{LTSmodel_cate}) also need to be estimated if a two-part model is used. Any appropriate statistical software for linear mixed model or generalized linear mixed model can be used for these $M$ steps.

For variance estimation, we propose to use the bootstrap. We first generated a bootstrap dataset by random sampling with replacement of the subjects in the training data. We then applied the MBJM estimation procedure above to obtain point estimators for the model parameters. We repeated this process 200 times and the sample variance of the point estimators is the bootstrap variance estimator. To calculate the variance of the prediction probability of a generic subject in the validation data, we can calculate this probability by using each of the 200 estimated model parameter sets from the bootstrap, and then estimate the sample variance of these estimated prediction probabilities. Wald-type confidence intervals can also be derived from the estimated variance and the asymptotic normality of the parameter estimators. Logit transformation of the prediction probability may be used to improve the finite-sample performance of the variance estimation procedure above, followed by a back transformation to obtain confidence intervals on the original probability scale. \citep{Li2005} As measures of prediction accuracy, we utilize the time-dependent AUC and time-dependent Brier score (BS). \citep{Li2016ROC, Wu2018} We utilized five-fold cross-validation to assess prediction accuracy and avoid overfitting bias in the real data analysis (Section \ref{sec:data.app}). The simulation studies used independently generated external validation datasets to estimate the prediction accuracy (Section \ref{sec:simu}).

\section{Theoretical Properties}
\label{sec:Theorem}

This section presents two theoretical results. The first one shows that the dynamic prediction model is expected to outperform the static prediction model, because the latter does not properly account for the longitudinal nature of this prediction problem. While this result is evidence for the importance of MBJM, it applies to all dynamic prediction models, including both landmark modeling and joint modeling approaches. The second one pertains to the asymptotic distribution of the CCA estimators of the MBJM.  

\begin{theorem}
	Let $\tilde{T}$ denote the event time and $\overline{\bY(\tT)} = \{ \bY(t); t \in (0, \tT) \}$ denote the history of longitudinal predictors up to $\tT$. The notation $\bY(t)$ covers time-varying and time-invariant predictors. The conditional distribution of survival data given the baseline predictor is $f_1(\tT | \bY(0); \balpha)$ with parameter $\balpha$. The joint distribution of longitudinal and survival data is $f_2(\tilde{T}, \overline{\bY(\tT)}; \bbeta)$ with parameter $\bbeta$. Suppose that these two distributions and the associated parameters are known.
	\newline
	For a subject who is at risk at the prediction time $s$, the static prediction model (SPM) defines the predicted probability of the outcome event by the horizon $\Delta$ as a function of the longitudinal predictors measured at $s$, calculated from the conditional distribution $f_1(.)$:
	\begin{equation}
		m_1( \bY(s); \balpha ) = P(\tT - s > \Delta | \tT > s, \bY(s); \balpha ) = \int_{\Delta}^{\infty} f_1( t | \bY(s) ; \balpha ) dt \in (0,1)  \notag   
	\end{equation}
	The DPM defines the predicted probability as a function of the longitudinal predictor history up to $s$, calculated from the joint distribution $f_2(.)$:
	\begin{equation}
		m_2(\overline{\bY(s)}; \bbeta) = P(\tT - s > \Delta | \tT > s, \overline{\bY(s)}; \bbeta ) \in (0,1)     \notag 
	\end{equation}
	Then, among subjects in the risk set $\tT > s$, the DPM has equal or better prediction accuracy than the SPM in terms of the Brier score, i.e., 
	\begin{equation}
    E\left\{ L_s\left( m_2(\overline{\mathbf{Y}(s)}; \boldsymbol{\beta}) \right) \mid \tT > s \right\} \leqslant E\left\{ L_s\left( m_1(\mathbf{Y}(s); \boldsymbol{\alpha}) \right) \mid \tT > s \right\}
\end{equation}
	where $L_{s}(P) = ( 1\{ \tT - s > \Delta \} - P)^2$ is the squared error loss for predicted probability $P$ at prediction time $s$.
	\label{thm1}
\end{theorem}

Theorem 1 shows that dynamic prediction is non-inferior to static prediction because it is adaptive to two longitudinal features of this prediction problem, time-varying at-risk population and time-varying longitudinal history. While this result is quite intuitive and has been discussed in previous literature, \citep{Li2016} the theorem expresses it in mathematical language. It is worthwhile to note that accounting for these two features does not always lead to better prediction. For example, in problems where the baseline time does not represent a clinical milestone in disease progression, the lack of adjustment for the at-risk population does not necessarily produce worse results. \citep{Yao2023} There could also be problems in which the longitudinal history of the biomarker does not further improve the prediction accuracy when the ``current'' biomarker value at the prediction time is incorporated in the model. Nonetheless, Theorem 1 shows that the prediction accuracy of dynamic prediction is at least not expected to perform worse than static prediction in these situations. 

According to the data notation in this paper, $T_i \leqslant \tau_{max}$ for all subjects. When $T_i = \tau_{max}$, $\delta_i = 0$ and $\tT_i > \tau_{max}$. When $T_i < \tau_{max}$, $\delta_i$ may equal to $0$ or $1$. The statement above holds when $\tT_i$ and $C_i$ are continuous variables, which implies that $P( T_i = \tau_{max} ) = 0$. When the time-to-event variables are ordinal or recorded in a real dataset with limited granularity, $T_i = \tau_{max}$ may occur but is uncommon. For MBJM-EX, we can set $\tau_{max} = \infty$. With this notation, the CCA of both MBJM-EX and MBJM-TP uses data from subjects with $\Delta_i = 1$, where  $\Delta_i = 1-(1-\delta_i) \boldsymbol 1\{T_i < \tau_{max}\} $. We call these subjects the ``unified group'' in the discussion below. The following result shows that a likelihood analysis of longitudinal measurements from this subset of study data yields consistent and asymptotically normal estimators for the longitudinal sub-model.  

\begin{theorem}
	Suppose $\bTheta_0$ is the true parameter of conditional probability $P(\bY|\tilde{T}, \bV; \bTheta)$, where $\bY =  ( \bm{Y}_{1}^T, \bm{Y}_{2}^T,...,\bm{Y}_{M}^T )^T $ is vector of longitudinal measurements of $M$ biomarkers. Let $l_i(\bTheta)=\log f(\bY_i | T_i, \delta_i, \bV_i; \bTheta)$ denote the log-likelihood function for observation $i$. Suppose $\hat{\bTheta}$ is the maximum likelihood estimator of the unified group using the likelihood function $\sum^n_{i=1} \Delta_i l_i(\bTheta)$, which is also a solution of the score equation: $\boldsymbol U_n(\bTheta)=\sum^n_{i=1} \frac{\partial}{\partial \bTheta} \Delta_i \log f(\bY_i|\tilde{T_i}, \bV_i;\bTheta)=0$. Then under some regularity conditions (see the online Supplementary Materials):
    \begin{enumerate}
        \item $\hat{\bTheta}$ is a consistent estimator of $\bTheta_0$;
         \item $\hat{\bTheta} \sim N \bigl(\bTheta_0, A(\bTheta_0)^{-1}B(\bTheta_0) \{A(\bTheta_0)^{-1}\}^{T}\bigr)$ as $n \rightarrow \infty$ where $A(\bTheta_0) = E\left\{ -\frac{\partial}{\partial \bTheta^T}U_n(\bTheta)|_{\bTheta = \bTheta_0} \right\}$ and $B(\bTheta_0) = E\left\{ U_n(\bTheta_0)U_n(\bTheta_0)^T \right\}$.
    \end{enumerate}
\end{theorem}

\section{Data Application}
\label{sec:data.app}
We evaluated the performance of the proposed method using a real dataset from a Primary Biliary Cirrhosis (PBC) study of the liver. The dataset is publicly accessible through the \texttt{joineRML} and \texttt{JMBayes2} packages of R. The dataset comprises 312 patients with a rare autoimmune liver disease, of whom 169 (54\%) reached the terminal composite event of death or liver transplant, whichever occurred first, and the rest were censored. Our primary objective is to dynamically predict this terminal composite event using baseline covariates and multiple biomarkers that were measured longitudinally during study visits. Our analysis used seven longitudinal biomarkers: bilirubin, prothrombin, albumin, alkaline, SGOT (serum glutamic-oxaloacetic transaminase), ascites, and hepatomegaly. It is important to note that hepatomegaly and ascites are categorical biomarkers (both are dichotomous), and the others are continuous variables. The baseline covariates include age and gender.

\begin{figure}[H]
    \centering
    \includegraphics[width = 0.7\textwidth]{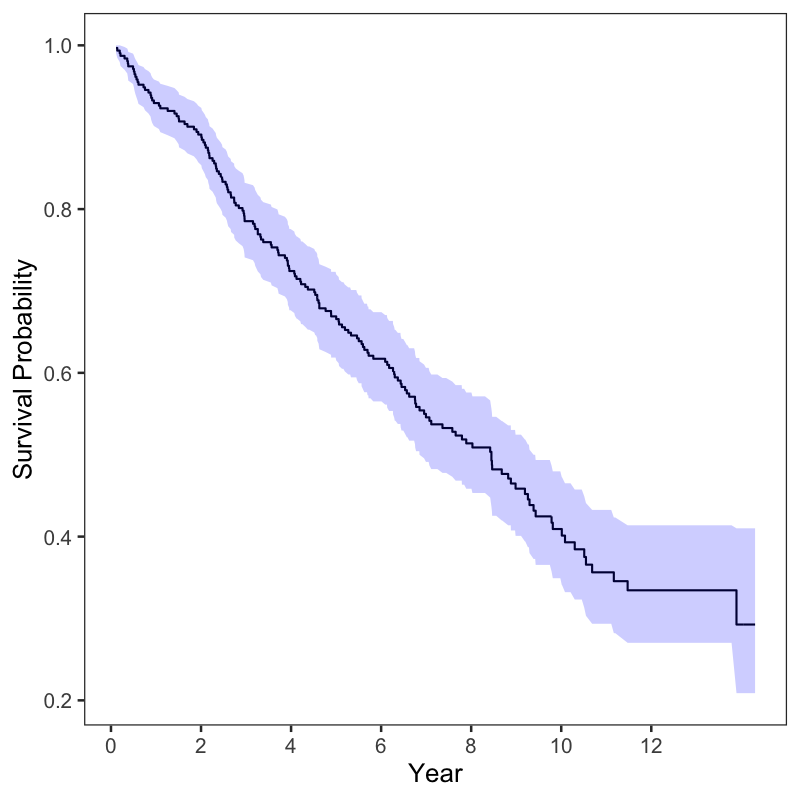}
    \caption{ \textit{The Kaplan-Meier survival curve of the terminal composite event in the PBC study.}}  \label{KM_PBC}
\end{figure}

\begin{figure}[H]
    \centering
    \includegraphics[width = 0.75\textwidth]{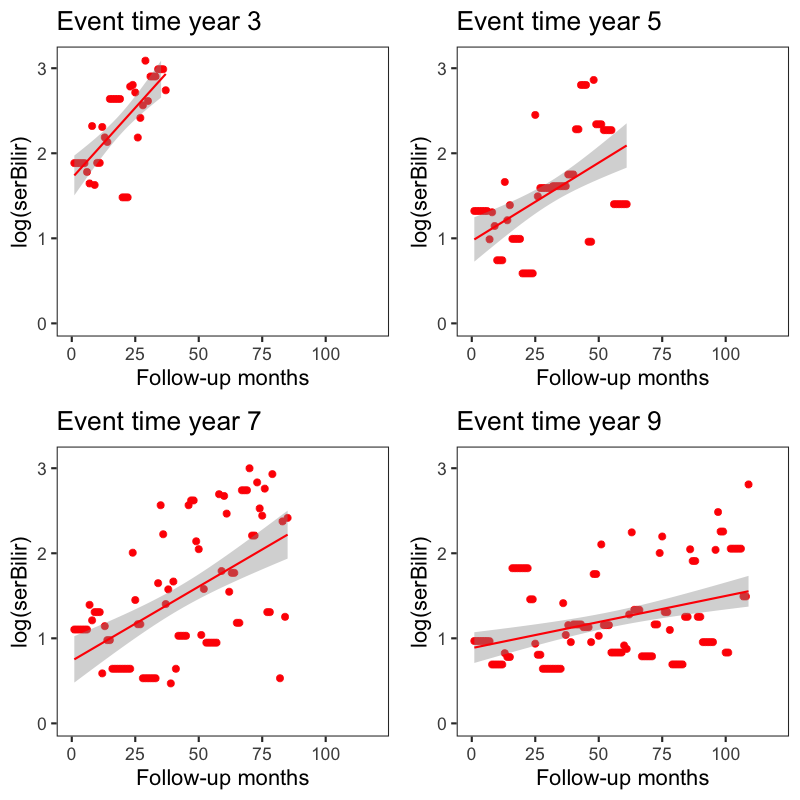}
    \caption{ \textit{The conditional mean trajectory (CMT) of the logarithm of serum bilirubin, $\log$(serBilir), at 4 selected strata of survival time in the PBC data analysis. Each plot was created by aggregating the longitudinal biomarker data from all participants who experienced the terminal event in that year. The red dots represent the average biomarker data by month. When the biomarker measurement times are not informative, these are unbiased estimators of the CMT at each month. A linear model (red line) with pointwise confidence intervals (gray band) estimated from the red dots data demonstrates an increasing trend within every survival strata, though the slope decreases with longer survival times. } } \label{cmt_serBilir}
\end{figure}

\begin{figure}[H]
    \centering
    \includegraphics[width = 1\textwidth]{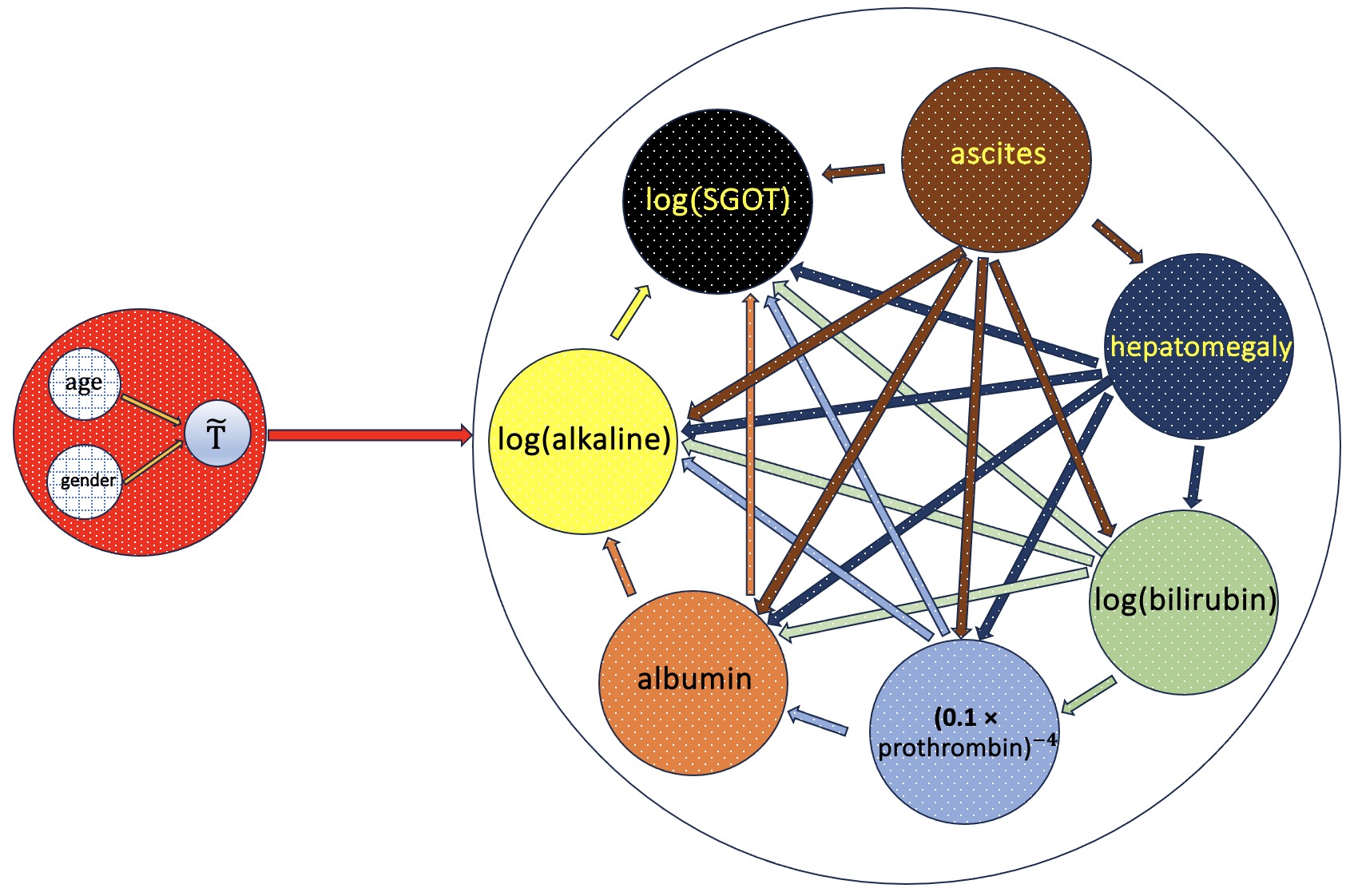}
    \caption{ \textit{A graphical representation of the MBJM used in the PBC analysis. The plotting symbols resemble those in Figure \ref{MBJM_model}, where an arrow that goes from A to B indicates that A is one of the covariates and B is the outcome in a regression model. The red circle on the left represents layer 0, the survival sub-model, and all the variables in the red circle serve as covariates in other layers. Layer 1 has ascites as the categorical outcome, Layer 2 has hepatomegaly as the categorical outcome, and Layers 3-7 have bilirubin, prothrombin, albumin, alkaline, and SGOT as the continuous outcomes.}}  \label{MBJM_PBC_2}
\end{figure}

Figure \ref{KM_PBC} shows the Kaplan-Meier curve of the terminal event that we want to predict, calculated from the training data. The curve descends to about 30\%, suggesting that most but not all the distribution of the terminal event is identifiable. Hence, BJM with extrapolation or two-part specification are both plausible, though the latter is more flexible at the cost of more model parameters. The follow-up extends to about 12 years, suggesting the dynamic prediction can be reliably made and evaluated in cross-validation before year $12 - \Delta$, where $\Delta$ is the prediction horizon, taken to be either 1 or 3 in our analysis (e.g., Table \ref{PBC_DP_Accuracy}). Figure \ref{cmt_serBilir} exhibits the conditional mean trajectories (CMTs) of log-transformed bilirubin, demonstrating an increasing trend in all survival strata but the rate of increase varies with the survival time. The CMTs of additional continuous biomarkers are displayed in Figures S1-S4 in the online Supplementary Materials. This is direct evidence that the longitudinal sub-model of bilirubin must incorporate the survival time as a covariate, and it is also indirect evidence that the longitudinal trajectory of bilirubin is predictive of survival. CMT plots of other longitudinal biomarkers are omitted for brevity. We designate $\tau_{max} = 12$ years for the MBJM-TP. Approximately 8\% of patients were followed up beyond $\tau_{max}$. The structure of the MBJM-TP is depicted in Figure \ref{MBJM_PBC_2}, resembling the model shown in Figure \ref{MBJM_model}. 

The survival sub-model is a Cox model with baseline covariates. The longitudinal sub-model for each of the 7 layers uses a random intercept to account for intrasubject correlation. We conducted a sensitivity analysis to study whether deviations from this basic model specification affect prediction accuracy (Table \ref{PBC_DP_sensitivity}). The deviations considered include adding random slopes to the model and adding a monotone transformation $g(\tT) = \log( \tT )$ to reduce the skewness of $\tT$ before entering it into the longitudinal sub-model. Table \ref{PBC_DP_sensitivity} shows that these additional features did not improve the prediction accuracy, and our final model retains the simplest specification without random slope or transformation of the survival time. The 7 longitudinal biomarkers enter the layers in the order shown in Figure \ref{MBJM_PBC_2}. We studied whether this order has any notable effect on the prediction accuracy by comparing it with the results from three alternative orders (Table \ref{PBC_DP_sensitivity}). In Order1, we alter the order among continuous biomarkers: SGOT, bilirubin, prothrombin, albumin, and alkaline in layers 3-7 respectively. In Order2, we alter the order of the two categorical biomarkers from Figure \ref{MBJM_PBC_2}. In Order3, we change the order of both continuous biomarkers (using Order1) and categorical biomarkers (using Order2). Table \ref{PBC_DP_sensitivity} shows that the prediction accuracy is not sensitive to the order of the biomarkers in the layers. Details of the estimated model can be found in the online Supplementary Materials.

We compared the prediction accuracy between the proposed MBJM and the shared random effects joint model (SJM), implemented by R package \texttt{JMBayes2} \citep{JMBayes2}. The SJM has been extensively studied \citep{Rizopoulos2012, Elashoff2016}. It uses linear mixed models for the longitudinal biomarkers and generalized linear mixed models for the categorical biomarkers, and these models share the same random effects with the model for survival data to build a joint distribution. The SJM is widely known to be computationally difficult with many longitudinal variables, and specialized software is needed to fit such models. The PBC data application requires the SJM software to accommodate multivariate longitudinal continuous and categorical variables. To our knowledge, such software was unavailable until a recent update to the \texttt{JMBayes2} package. Table \ref{PBC_DP_Accuracy} shows that MBJM-EX, MBJM-TP, and the SJM model are comparable in terms of both the AUC and the Brier score. However, the computation time required for MBJM is significantly less than that for SJM. Specifically, MBJM-EX takes approximately 111 seconds and MBJM-TP takes 85 seconds to complete the five-fold cross-validation on a personal computer with 2.9 GHz CPU and 32 GB RAM, whereas SJM requires 76 minutes. Section \ref{sec:simu} presents a more extensive comparison of the computing speed between MBJM and SJM. It is also worth noting that the MBJM can be implemented with standard software for linear mixed models, generalized linear mixed models, and survival models, without needing customized packages. The standard software is either commercially available or open-source but with active maintenance; the customized software is an important supplement, but the maintenance critically depends on the owner and the invested resources. Table \ref{PBC_DP_Accuracy} also includes a comparison between MBJM, SJM and SPM. The dynamic prediction models have equal or better prediction accuracy than the static prediction model, as suggested by the theoretical result in Section \ref{sec:Theorem}.

\begin{figure}[H]
    \centering
    \includegraphics[width=1.0\textwidth]{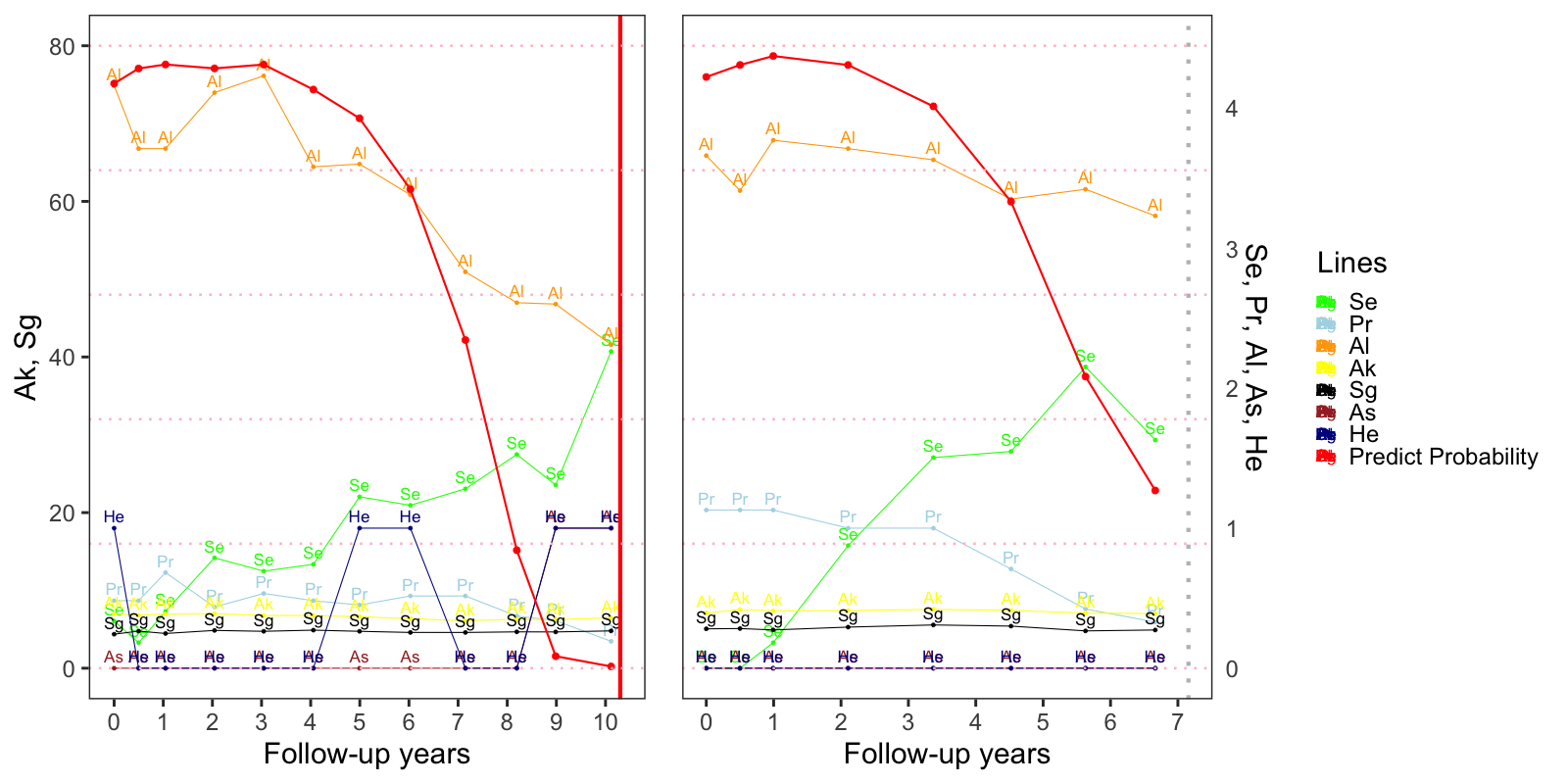}
    \caption{ \textit{Illustration of the dynamic prediction of the terminal event outcome with two patients from the PBC data, patient A on the left and patient B on the right. The horizontal axis is the time since baseline. The individual trajectories of the seven longitudinal biomarkers are plotted in different colors. The trajectories are formed by connecting the repeated measures with line segments. The continuous biomarkers of alkaline phosphatase (Ak) (U/liter) and SGOT (Sg) (U/ml) are plotted to the vertical axis on the left. Both Ak and Sg are transformed using $\log$. The continuous biomarkers of serum bilirubin (Se) (mg/dl), prothrombin time in seconds (Pr), and albumin (Al) (gm/dL) are plotted to the vertical axis on the right. Serum bilirubin is transformed with $\log$ and prothrombin is transformed with $(0.1 * \text{prothrombin})^{-4}$. Two categorical longitudinal biomarkers, ascites (As) and hepatomegaly (He), are plotted to the vertical axis on the right; they are dichotomous, taking a value of either 0 or 1. The six horizontal dotted lines in the background indicate the probability values of 0, 0.2, 0.4, 0.6, 0.8, and 1.0, from bottom to top. Predictions are made at the clinical visits when the biomarkers are measured. The predicted probabilities of surviving beyond a three-year horizon are shown as red dots, connected by linear segments to form a continuous curve. The vertical red line on the left plot marks the terminal event time for patient A, while the vertical grey dotted line on the right plot marks the censoring time for patient B. These dynamic prediction plots visualize how the personalized, real-time risks vary with the seven longitudinal biomarkers. The baseline predictors have been incorporated into the prediction. } } \label{PBC_individual_risk}
\end{figure}

Figure \ref{PBC_individual_risk} illustrates the dynamic prediction for two PBC patients. Patient A experiences a notable increase in serum bilirubin. The PBC impairs the liver's ability to process and excrete bilirubin. Elevated levels of serum bilirubin often signify disease progression, which can lead to liver transplant or death. The predicted probability of survival sharply declines, and patient A reached the terminal endpoint at about year 10. Patient B exhibited a similar pattern of serum bilirubin as patient A, but the other biomarkers of liver conditions, such as the categorical biomarker hepatomegaly, suggest that this patient has better liver conditions than patient A. Of note, the prediction horizon is 3 years. Patient A's predicted probability of survival beyond the next 3 years dropped to very close to 0 by year 8 and later, and this patient experienced the terminal event at year 10. Patient B was censored at year 7; the prognosis at the last clinical visit was not as bad as the prognosis of patient A at the last two visits.

\section{Simulation}
\label{sec:simu}

We conducted two simulation experiments to study the performance of our proposed model and compared it with the SJM and SPM. The first experiment uses data that were simulated from either the MBJM-EX or MBJM-TP and the goal is to demonstrate the correct estimation of the model parameters by the proposed CCA procedure when the model assumptions are satisfied. The second experiment compares the predictive performance of MBJM and SJM. Since these two types of joint models have their respective modeling assumptions and cannot be specified correctly simultaneously, we simulated data from both MBJM and SJM for a fair comparison. We also compared MBJM and SJM with SPM to demonstrate the advantage of dynamic prediction over static prediction, as suggested by the theoretical results in Section \ref{sec:Theorem}. 

In all simulation experiments, we generated datasets that closely resemble the characteristics of the PBC data. We used 7 longitudinal biomarkers, with 5 continuous and 2 categorical, as in the PBC data. To our knowledge, such a large number of longitudinal variables have rarely been used in the published literature of joint modeling of longitudinal and survival data. For each simulation scenario, we used 300 Monte Carlo repetitions unless stated otherwise, and the results were aggregated. For a prediction accuracy assessment without overfitting bias, each Monte Carlo repetition includes a training dataset and a validation dataset with equal sample size, both simulated from the same distribution. The training dataset has the pre-specified censoring rate, but the validation dataset contains no censoring in order to calculate the prediction accuracy measures with better precision and robustness. The numerical details of the simulation setting and data generation procedures are in the online Supplementary Materials.   

\begin{figure}[H]
    \centering
    \includegraphics[width = 1.1\textwidth]{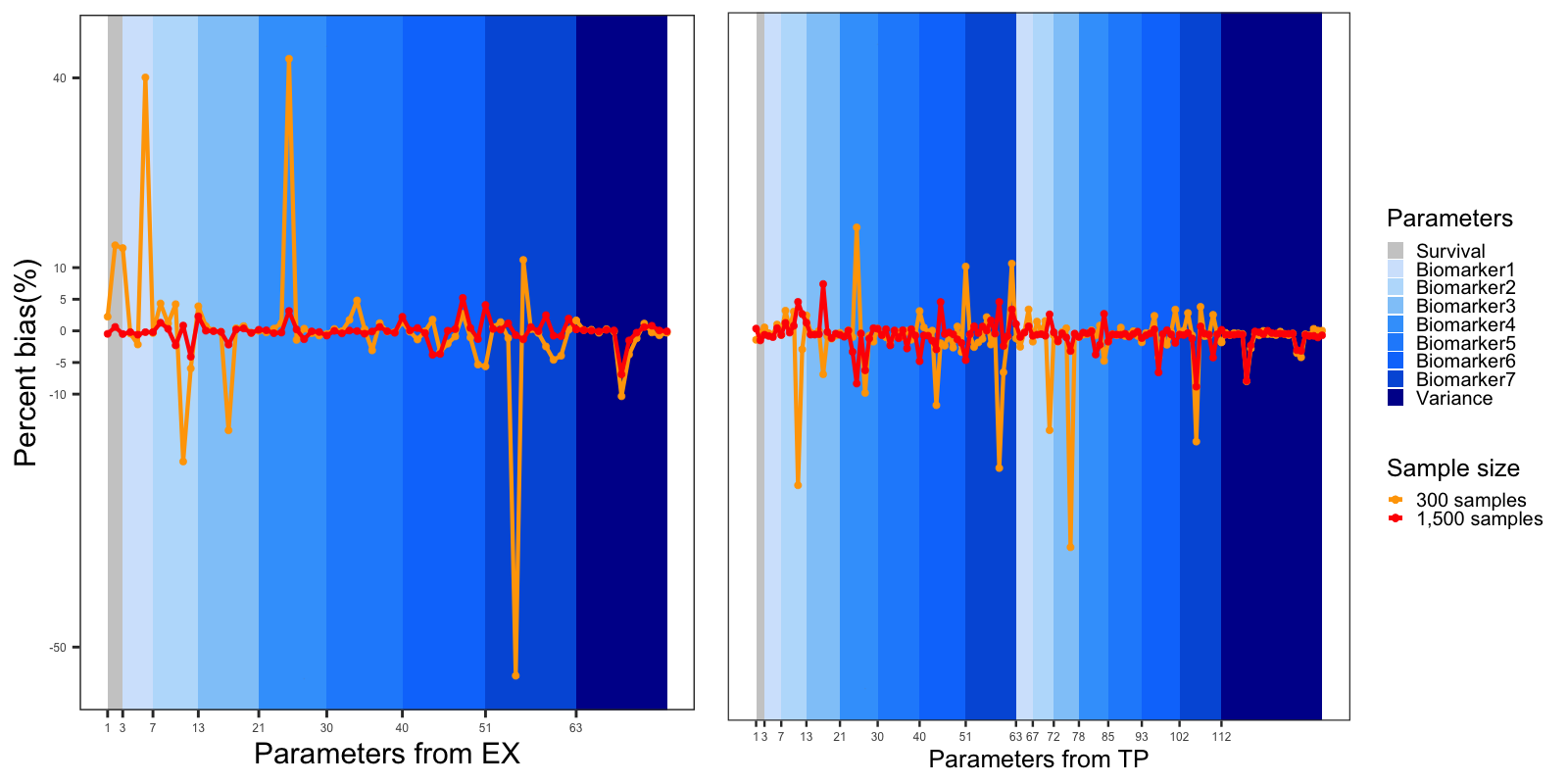}
       \caption{ \textit{The bias of the estimated model parameters from the simulation study. The plot on the left shows results with the MBJM-EX model. The plot on the right shows results with the MBJM-TP model. The horizontal axis represents the index of parameters, which are grouped based on their roles in the model. To enhance the visual representation, distinct background colors have been assigned to different groups of parameters. In the EX model, the parameter groups are arranged from left to right as follows: the survival sub-model, longitudinal sub-models for biomarkers 1-7, and the variance-covariance matrix. In the TP model, the parameter groups are arranged from left to right as follows: the survival sub-model, longitudinal sub-models for biomarkers 1-7, longitudinal LTS sub-models for biomarkers 1-7, and the variance-covariance matrix. The number of parameters in each group can be determined by referring to the horizontal axis. The two curves in the figure represent the percent bias, which is plotted to the vertical axis on the left. The orange curve corresponds to a sample size of 300, while the red curve represents 1,500.  } }    \label{Percent_bias}
\end{figure}

We first studied the consistency of parameter estimation of both MBJM-EX and MBJM-TP, under a sample size of either 300 or 1,500. The data were simulated under the respective modeling assumptions of these two models. Two baseline covariates were simulated to mimic age and gender distribution in the PBC data. The 5 continuous and 2 categorical longitudinal biomarkers were simulated from either linear mixed or generalized linear mixed models. Survival data were simulated from a Weibull distribution. The censoring rate for MBJM-EX was approximately 45\%. For MBJM-TP, the overall censoring rate was 45\%, with 15\% of subjects being censored at $\tau_{max}$. Figure \ref{Percent_bias} illustrates the finite-sample biases for MBJM-EX and MBJM-TP. As the sample size grows from 300 to 1,500, the percent bias decreases to almost zero. This is evidence of consistent estimation. The mean squared errors also reduced substantially, though that result was omitted here for brevity.  

Next, we compared the prediction accuracy between MBJM and SJM. The SJM was implemented similarly as in the PBC data analysis. Since we found similar prediction accuracy between MBJM-EX and MBJM-TP in the PBC data analysis, we only used MBJM-EX in this comparison. There are two scenarios: Sim-MBJM, where the data were simulated from MBJM-EX, and Sim-SJM, where the data were simulated from SJM. In both scenarios, we chose a numerical setting that is close to the fitted model to the PBC data (for MBJM-EX, the setting is identical to the simulation in Figure \ref{Percent_bias}). Since it is time-consuming to estimate the SJM, we reduced the number of Monte Carlo repetitions to 100. The training dataset has 500 subjects, with a censoring rate of approximately 45\% under both Sim-MBJM and Sim-SJM scenarios. 

\begin{figure}[H]
    \centering
    \includegraphics[width = 0.7\textwidth]{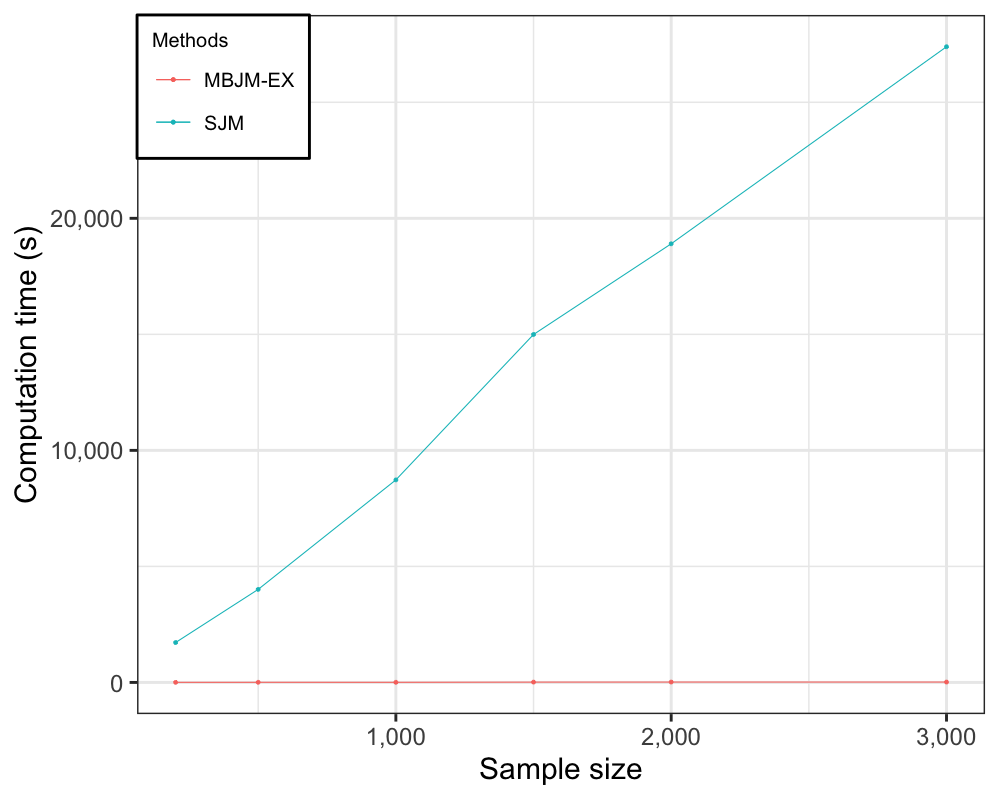}
    \caption{ \textit{Computing times of the MBJM-EX and SJM in the model fitting with a single simulated dataset. The time is expressed as seconds. The computation was done on a personal computer with a 2.9 GHz CPU and 32 GB RAM. The sample sizes of the simulated datasets vary at 200, 500, 1000, 1500, 2000, and 3000. }} \label{time_compare}
\end{figure}

Table \ref{Sim_DP_Accuracy} shows that the predictive accuracy of the MBJM is comparable to the SJM, irrespective of the data generation scenarios, and their performances are both notably better than the static prediction, as predicted by the theoretical result in Section \ref{sec:Theorem}. Both MBJM and SJM demonstrated some robustness against model misspecification in this simulation experiment, partially attributed to their flexibility. However, the MBJM model can be implemented with standard regression software for survival data and repeated measures data, and requires substantially less computing time compared to the SJM. We conducted a small-scale simulation experiment with sample sizes of 200, 500, 1000, 1500, 2000, and 3000. Under each sample size, five Monte Carlo repetitions are performed and the average computation time is recorded. Figure \ref{time_compare} shows large time-savings in computing times with the MBJM, especially when the sample size is large. Moreover, as the sample size approached 3000, the computation time for analyzing a single dataset by SJM increased to approximately 8 hours, and it achieved convergence in only two out of the five Monte Carlo repetitions. In contrast, the MBJM only used 15.95 seconds and had a 100\% convergence rate. 


\section{Discussion}
\label{sec:disc}
This paper extends the BJM methodology for dynamic prediction \citep{Shen2021} to the MBJM, which is designed to accommodate a mixture of continuous and categorical longitudinal variables. The extension uses an alternative layered model formulation to avoid possible high-dimensional integration from the model of longitudinal categorical data. The MBJM can be estimated conveniently by a complete-case analysis with standard software, and the computation is fast and robust. Previous research in dynamic prediction suggests that the landmark modeling approach is computationally fast and easier to implement with standard software, but has difficulty fully accounting for the longitudinal history \citep{Li2016}; the joint modeling, on the other hand, is computationally more intensive and must be implemented with customized software, but it is straightforward to account for longitudinal history. The methodology in this paper brings the joint modeling to the same level as the landmark modeling in computational tractability while retaining the joint modeling's fully likelihood-based prediction calculation. The theoretical and empirical comparison with the static prediction model further demonstrates the usefulness of the proposed methodology, though the scope of application extends beyond the MBJM.   

The MBJM utilizes a large number of fixed effect parameters (i.e., population-averaged parameters) to model the correlation among longitudinal variables. This is perhaps inevitable in a flexible model of the joint distribution of longitudinal and survival data. In the SJM, for example, such correlations are typically modeled by many random effects, which results in a large variance-covariance matrix with the number of unknown parameters increasing quadratically with the number of random effects. While the MBJM has more fixed effect parameters than the SJM, it has substantially fewer random effect parameters (i.e., subject-specific parameters). The fixed effect models are generally easier to estimate and model diagnosis procedures are also better developed. Similar to the parameters in the variance-covariance matrix of random effects in an SJM, the fixed effect parameters in the MBJM are nuisance parameters for prediction purposes; their interpretation is of secondary importance. From this perspective, the formulation of MBJM resembles some machine learning methods and sacrifices interpretability for convenience in prediction.

The prediction accuracy of a joint model may deteriorate under model misspecification. \citep{Li2023} The prediction equation (\ref{DP_risk}) and the likelihood factorization (\ref{eq:series.conditional}) cannot be misspecified. The only possible source of model misspecification is the linear or generalized linear mixed model for the longitudinal data layers, and the survival model with baseline covariates. However, these are standard regression models with decades of development, numerous extensions, and abundant software. The model misspecification can be significantly mitigated, as demonstrated in our simulation study. The MBJM can be viewed as an analytical framework with many different model formulations, all of which centered around (\ref{DP_risk}) and (\ref{eq:series.conditional}), rather than a specific joint model.  

A limitation of this study is that all the biomarkers should be measured at the same series of visit times of a subject, i.e., $t_{mij} = t_{ij}$ for all $m=1,2,..., M$, though different subjects are allowed to have different visit times. This assumption may not hold in some situations. For example, expensive or invasive biomarker tests may be performed less frequently than routine clinical tests. This issue warrants further research in the future. The conventional censoring assumption in the joint modeling literature, as described in Section \ref{sec:estimand_model}, can be relaxed by adding a sub-model for the censoring time. Diggle et al \cite{Handbook.Longitudinal.Chap15} provided an example of how this can be done within the framework of the shared random effects models, though this is an uncommon practice in the joint modeling literature. It is of future research interest to study how the sub-model for the censoring time should be designed and incorporated into the BJM framework to allow dependence between the censoring time and longitudinal and terminal event data.

\section*{Funding}
This research is supported by the National Institutes of Health [grant numbers P30CA016672 to L.L., R01DK118079 to W.L. and L.L., R01HL175410 to Z.Y. and L.L.]. 

\bibliographystyle{vancouver}
\bibliography{MBJMRef}

\newpage
\begin{table}[H]
	\centering
		\caption{Evaluating the sensitivity of prediction accuracy results to MBJM specification in the PBC data analysis. MBJM-EX: MBJMs with extrapolation. RI: the longitudinal sub-model includes random intercepts. RS: the longitudinal sub-model includes both random intercepts and slopes. Log: the longitudinal sub-model uses $g(x) = \log(x)$ to reduce the skewness of $\tT$. MBJM-EX-RI places the 7 longitudinal biomarkers in the layers according to the order in Figure \ref{MBJM_PBC_2}. We changed the ordering of continuous biomarkers only (Order1), categorical biomarkers only (Order2), and both (Order3). The prediction accuracy was evaluated for at-risk subjects at Year 1, 3, and 5 after baseline. The prediction horizons were 1 and 3 years. All prediction accuracy results were from cross-validation. This sensitivity analysis did not use the two-part MBJM because its performance was found similar to MBJM-EX in Table \ref{PBC_DP_Accuracy}. }\label{PBC_DP_sensitivity}
		\begin{tabular}{ccccccccccccccc}
		\hline
		Accuracy & Horizon& Model&  Year1 & Year3  &Year5   \\ 
		\hline
  		\multirow{14}{*}{AUC}  &\multirow{7}{*}{1}& MBJM-EX-RI   &0.8331 & 0.8530 & 0.8407 \\
		&& MBJM-EX-RS   &0.8370& 0.8556 &0.8354  \\
		&  &MBJM-EX-RI-Log&0.8367 & 0.8551 & 0.8314\\ 
  		&  &MBJM-EX-RS-Log&0.8397 &0.8616 &0.8247\\ 
        &  &MBJM-EX-RI-Order1&0.8321 &0.8548 &0.8348\\ 
        &  &MBJM-EX-RI-Order2&0.8344 &0.8530 &0.8417\\ 
        &  &MBJM-EX-RI-Order3&0.8340 &0.8543 &0.8349\\ 

        \cline{2-6}
      
        &\multirow{7}{*}{3}& MBJM-EX-RI   & 0.9014 & 0.8643 & 0.8233  \\
		&& MBJM-EX-RS   &0.9048 &0.8664 &0.8082   \\
		&  &MBJM-EX-RI-Log & 0.8982 &0.8589 & 0.8483  \\ 
  		&  &MBJM-EX-RS-Log & 0.9017 & 0.8603 & 0.8271  \\ 
        &  &MBJM-EX-RI-Order1 & 0.9056 &0.8633 &0.8224\\
        &  &MBJM-EX-RI-Order2 & 0.9007 &0.8639 &0.8241\\
        &  &MBJM-EX-RI-Order3 & 0.9058 &0.8634 &0.8231\\
		\hline

   		\multirow{14}{*}{BS}&\multirow{7}{*}{1} & MBJM-EX-RI & 0.0336& 0.0551 &0.0678\\ 
  		& & MBJM-EX-RS & 0.0338 &0.0551& 0.0673   \\ 
		& &  MBJM-EX-RI-Log & 0.0358 &0.0578 &0.0624\\ 
  		& &  MBJM-EX-RS-Log & 0.0360 & 0.0576 & 0.0600\\ 
        & &MBJM-EX-RI-Order1 & 0.0342 & 0.0547 & 0.0680\\ 
        & &MBJM-EX-RI-Order2 & 0.0335 &0.0552 &0.0677\\ 
        & &MBJM-EX-RI-Order3 & 0.0341 &0.0548 &0.0680\\ 

        \cline{2-6}
            
        &\multirow{7}{*}{3} & MBJM-EX-RI & 0.0969 & 0.1164 &0.1614 \\ 
  		& & MBJM-EX-RS &  0.0956 &0.1150& 0.1664 \\ 
		& &  MBJM-EX-RI-Log &0.1027 &0.1178 &0.1314 \\
   		& &  MBJM-EX-RS-Log &0.1005 &0.1157 &0.1333 \\
        & &MBJM-EX-RI-Order1 & 0.0943 &0.1163 &0.1589\\
        & &MBJM-EX-RI-Order2 & 0.0971 & 0.1165 & 0.1610\\
        & &MBJM-EX-RI-Order3 & 0.0946 & 0.1165 & 0.1586\\
        
	    \hline
		\multicolumn{3}{c}{\# at-risk patients in PBC data}& 290& 245& 200\\
		\hline
	\end{tabular}
	
\end{table}

\newpage
\begin{table}[H]
	\centering
		\caption{Prediction accuracy results of MBJM-EX, MBJM-TP, SJM and SPM in the PBC data analysis. SJM: shared random effects joint model. SPM: static prediction model. The prediction accuracy was evaluated for at-risk subjects at Year 1, 3, and 5 after baseline. The prediction horizons were 1 and 3 years. All prediction accuracy results were from cross-validation.  }\label{PBC_DP_Accuracy}
		\begin{tabular}{ccccccccccccccc}
		\hline
		Accuracy & Horizon& Model&  Year1 & Year3  &Year5   \\ 
		\hline
  		\multirow{6}{*}{AUC}  &1& MBJM-EX   &0.8331 & 0.8530 & 0.8407 \\
		   &1& MBJM-TP   &0.8331 & 0.8563 & 0.8257  \\
		   & 1 &SJM& 0.7905 & 0.8653 & 0.8446\\ 
  		   & 1 &SPM& 0.7974 &0.8919 &0.7172\\
    
           &3& MBJM-EX   & 0.9014 & 0.8643 & 0.8233  \\
		   &3& MBJM-TP   &0.9025 &0.8620 &0.8416   \\
		   & 3 &SJM & 0.8849 & 0.8790 & 0.8354  \\ 
     & 3 &SPM& 0.9009& 0.8520 &0.8076 \\
		\hline

   		\multirow{6}{*}{BS}&1 & MBJM-EX& 0.0336 &0.0551 &0.0678 \\ 
  		 &1 & MBJM-TP& 0.0348& 0.0545 &0.0695 \\ 
		 &1 &  SJM&0.0381 & 0.0601 & 0.0565\\ 

         &1 &  SPM&0.0414 &0.0535 &0.0606\\ 
         &3 & MBJM-EX& 0.0969 & 0.1164 &0.1614 \\ 
  		 &3 & MBJM-TP&  0.0951 & 0.1145 & 0.1553 \\ 
		 &3 &  SJM &0.1042 & 0.1121 & 0.1199 \\ &3 &  SPM &0.1055 &0.1162& 0.1246 \\ 

	    \hline
		\multicolumn{3}{c}{\# at-risk patients in PBC data}& 290& 245& 200\\
		\hline
	\end{tabular}
	
\end{table}

\clearpage
\begin{table}[H]
	\centering
		\caption{Prediction accuracy results of MBJM, SJM and SPM in the simulation. SJM: shared random effects joint model. SPM: static prediction model. Sim-MBJM: data simulated from MBJM. Sim-SJM: data simulated from SJM. The prediction accuracy was evaluated among at-risk subjects at Year 1, 3, and 5 after baseline. The prediction horizons are 1 and 3 years.  }\label{Sim_DP_Accuracy}
		\begin{tabular}{ccccccccccccccc}
		\hline
		Data & Accuracy & Horizon& Model&  Year1 & Year3  &Year5   \\ 
		\hline
  	    \multirow{8}{*}{Sim-MBJM }&   	    \multirow{4}{*}{AUC }  &1& MBJM-EX   &0.9455 & 0.9527& 0.9554 \\
		&    & 1 &SJM     & 0.9386 & 0.9455&0.9481\\ 
    	&    & 1 &SPM     & 0.7443 &0.7877 & 0.8350\\ 
  		
          &    &3& MBJM-EX   &0.9620 &0.9693 &0.9747    \\
		  &    & 3 &SJM & 0.9564 & 0.9631& 0.9696  \\ 
    	 &    & 3 &SPM & 0.7858 &0.8327 &0.8788  \\ 
		\cline{2-7}

   	      & \multirow{4}{*}{BS }&1 & MBJM-EX& 0.0537 &0.0501& 0.0558   \\ 
		  &  &1 &  SJM&0.0417 & 0.0439 &0.0435\\ 
    	 &  &1 &  SPM&0.1193& 0.1590 &0.1973\\ 
	  	
         &  &3 & MBJM-EX& 0.0644& 0.0614 &0.0555 \\ 
		&  &3 &  SJM &0.0671 & 0.0631 &0.0585  \\ 	
    	&  &3 &  SPM &0.2222& 0.1762 &0.1426  \\ 

	    \cline{2-7}
		&\multicolumn{3}{c}{\# at-risk patients in simulation data}& 474 & 416 &358\\
		\hline
    	\multirow{8}{*}{Sim-SJM} & \multirow{4}{*}{AUC}  &1& MBJM-EX   &0.8968& 0.9274& 0.9298\\
		  &    & 1 &SJM     &0.8977 & 0.9243& 0.9271\\ 
    	&    & 1 &SPM     &0.9449 &0.9528 &0.9570\\ 
  		
          &    &3& MBJM-EX   &0.9106 & 0.9433 & 0.9427    \\
		  &    & 3 &SJM &0.9099& 0.9434 &0.9396  \\ 
    	 &    & 3 &SPM &0.9528 &0.9570 &0.9433  \\ 
		\cline{2-7}

   	     & \multirow{4}{*}{BS}&1 & MBJM-EX& 0.0298 &0.0443& 0.0597   \\ 
		  &  &1 &  SJM& 0.0345 & 0.0508 &0.0614\\ 
    	&  &1 &  SPM& 0.0619 &0.1418 &0.2754\\ 
	  	
          &  &3 & MBJM-EX&0.0797& 0.0802 & 0.0909 \\ 
		  &  &3 &  SJM & 0.0797& 0.0789 &0.0920 \\ 	
    	&  &3 &  SPM & 0.1031 &0.2146& 0.3750 \\ 

	    \cline{2-7}
		&\multicolumn{3}{c}{\# at-risk patients in simulation data}& 476 &439 &378\\
		\hline
	\end{tabular}
	
\end{table}

\newpage

\newpage

\newpage


\newpage

\newpage

\newpage

\end{document}